\documentclass[12pt,preprint]{aastex}
\begin{document}
\shorttitle{Type IIn Supernovae 2010jl}

\shortauthors{Zhang, T. et al.}

\title{Type IIn Supernova SN 2010jl: Optical Observations for Over 500 Days After Explosion}

\author{Tianmeng Zhang\altaffilmark{1,2}, Xiaofeng Wang\altaffilmark{3}, \\
Chao Wu\altaffilmark{1}, Juncheng Chen\altaffilmark{3}, Jia Chen\altaffilmark{3}, Qin Liu\altaffilmark{3}, Fang Huang\altaffilmark{3,4}, Jide Liang\altaffilmark{3}, Xulin Zhao\altaffilmark{3},\\ Lin Lin\altaffilmark{5}, Min Wang\altaffilmark{6}, Michel Dennefeld\altaffilmark{7}, Jujia Zhang\altaffilmark{8},\\ Meng Zhai\altaffilmark{1,2}, Hong Wu\altaffilmark{1,2}, Zhou Fan\altaffilmark{1,2}, Hu Zou\altaffilmark{1,2}, Xu Zhou\altaffilmark{1,2}, Jun Ma\altaffilmark{1,2}}

\altaffiltext{1}{National Astronomical Observatories of China, Chinese Academy of Sciences, Beijing 100012, P.R. China; armengjade@gmail.com}
\altaffiltext{2}{Key Laboratory of Optical Astronomy, National Astronomical Observatories, Chinese Academy of Sciences, Beijing 100012, P.R. China}
\altaffiltext{3}{Physics Department and Tsinghua Center for Astrophysics (THCA), Tsinghua University, Beijing 100084, China; wang\_xf@mail.tsinghua.edu.cn}
\altaffiltext{4}{Beijing Normal University, Beijing, 100875, P. R. China}
\altaffiltext{5}{University of Science and Technology of China, No.96, JinZhai Road Baohe District, Hefei, Anhui, 230026,P.R.China}
\altaffiltext{6}{Purple Mountain Observatory, Chinese Academy of Sciences, Nanjing, 210008, P.R. China}
\altaffiltext{7}{Institut d'Astrophysique de Paris, and University Pierre et Marie Curie (Paris 6)}
\altaffiltext{8}{Yunnan Astronomical Observatory, Chinese Academy of Sciences, Kunming 650011, P.R. China}

\begin{abstract}

We present extensive optical observations of a Type IIn supernova (SN) 2010jl for the first 1.5 years after the discovery. The $UBVRI$ light curves demonstrated an interesting two-stage evolution during the nebular phase, which almost flatten out after about 90 days from the optical maximum. SN 2010jl has one of the highest intrinsic H$\alpha$ luminosity ever recorded for a SN IIn, especially at late phase, suggesting a strong interaction of SN ejecta with the dense circumstellar material (CSM) ejected by the progenitor. This is also indicated by the remarkably strong Balmer lines persisting in the optical spectra. One interesting spectral evolution about SN 2010jl is the appearance of asymmetry of the Balmer lines. These lines can be well decomposed into a narrow component and an intermediate-width component. The intermediate-width component showed a steady increase in both strength and blueshift with time until t $\sim$ 400 days after maximum, but it became less blueshifted at t $\sim$ 500 days when the line profile appeared relatively symmetric again. Owing to that a pure reddening effect will lead to a sudden decline of the light curves and a progressive blueshift of the spectral lines, we therefore propose that the asymmetric profiles of H lines seen in SN 2010jl is unlikely due to the extinction by newly formed dust inside the ejecta, contrary to the explanation by some early studies. Based on a simple CSM-interaction model, we speculate that the progenitor of SN 2010jl may suffer a gigantic mass loss ($\sim$ 30-50 M$_{\odot}$) in a few decades before explosion. Considering a slow moving stellar wind (e.g., $\sim$ 28 km s$^{-1}$) inferred for the preexisting, dense CSM shell \citep{smith11b, and11} and the extremely high mass-loss rate (1-2 M$_{\odot}$ yr$^{-1}$), we suggest that the progenitor of SN 2010jl might have experienced a red supergiant stage and explode finally as a post-red supergiant star with an initial mass above 30-40 M$_{\odot}$.
\end{abstract}


\section{Introduction}

Type II supernovae (SNe II) have long been thought to arise from the death of massive stars with $M >$ 8~M$_{\odot}$ when the nuclear burning does not provide thermal pressure to support the star (e.g., \citet{smar09}). Observationally, the presence of hydrogen in the spectra of SNe II are the distinguishing signatures of this class (see \citet{fili97} for a review of SN classification), and they are usually fainter than the type Ia events. Type IIn supernovae (SNe IIn) represent a distinct subclass of SNe II that are characteristic of prominent narrow emission lines of hydrogen \citep{sch90, fili91a, fili91b}. They can also have intermediate-width and/or broader components in addition to the narrow component. The narrow component usually has a full-width-half-maximum (FWHM) velocity less than a few hundred km s$^{-1}$ and is believed to arise from the pre-shocked, photo-ionized circumstellar material (CSM) shell around the exploding star \citep{cf94}; while the intermediate- or broad-width component has an FWHM velocity of a few thousand km s$^{-1}$, and is formed from the dense post-shocked CSM shell. The presence of high-density CSM shells around some SNe IIn is favored by the detections of strong X-ray and radio emission after their explosion.

Type IIn supernovae are the brightest subgroup of all the SNe II, but they are rare events, accounting for $<$10\% of all the SNe II \citep{smar09, lwd11, smith11a}. The well-observed, representative SNe IIn sample are SNe 1988Z \citep{tur93}, 1994W \citep{soll98, chu04}, 1998S \citep{fas00, liu00, pool02}, 2006tf \citep{smith08}, and SN 2006gy \citep{smith07a}, which exhibit diverse observational properties. For example, the maximum-light absolute magnitude can vary from $\sim$ $-$19.0 mag (for SNe 1988Z and 1994W) to $\sim$ $-$22.0 mag (for SN 2006gy) in the broadband $R$. The strength and profile of the emission lines such as the H$\alpha$ lines also differ significantly among these SNe IIn \citep{smith08}. This indicates a large spread for the properties of the progenitor stars for this subgroup of core-collapse SNe (see a summary by \citet{kie12}).

The dense circumstellar medium can be formed due to a high mass-loss rate of massive stars \citep{chu04, gal09}. The luminous blue variable (LBV) stars are very bright, blue, hypergiant variable stars that experiences frequent eruptions \citep{hd94}, and has been proposed to be the possible progenitors of some type IIn SNe. One well-known case is the luminous type IIn SN 2005gl. Pre-explosion $Hubble$ $Space$ $Telescope$ (HST) images show that the bright point source at the position of SN 2005gl is likely an LBV-like star, and the speed of the pre-shocked progenitor wind is also consistent with an LBV star \citep{gal07, gal09}. Another case is SN 2006tf which was also suggested to have a pre-shocked wind speed of a LBV-like star, but not of red supergiants or Wolf-Rayet stars \citep{smith08}. On the other hand, it is also proposed that the WR star with fast winds lasting for a few thousand years might be the probable progenitors of some SNe IIn \citep{dvv10, dvv11}. Note that not all of the type IIn supernovae show strong radio or X-ray emission suggests that not all of them are undergoing a strong interaction with a dense CSM. This suggests that SNe IIn may have more than one type of progenitor.

SN 2010jl was discovered on November 02.06 UT \citep{np10} in an irregular galaxy UGC 5189A, and is one of the brightest supernova recorded in 2010 (13.5 mag on 2010 November 3). The early observation shows that it is a type IIn in the relatively young phase, with distinguished signatures of multiple hydrogen emission lines in the spectrum \citep{ben10}. At the optical position of the SN, an X-ray point-source was also detected on November 5 with a 6.5-$\sigma$ significance of source detection \citep{imm10}. Using the archival WFPC2 images of $Hubble$ $Space$ $Telescope$ taken at about 10 years prior to explosion, \citet{smith11b} suggests that the progenitor of SN 2010jl be consistent with a massive star with an initial mass above 30~M$_{\odot}$ but the exact nature of the progenitor is not conclusive. The optical linear spectropolarimetry of SN 2010jl obtained at two weeks after the discovery shows a continuum polarization at a level of 1.7-2.0\%, indicating that the explosion has a substantial asphericity \citep{pat11}. Near infrared observations were obtained at about 90 days after the explosion with the $Wisconsin$ $Indiana$ $Yale$ $NAOA$ telescope for the $JHK_{s}$ bands and the $Spitzer$ IRAC for two mid-infrared bands (with the wavelength centering at 3.6 and 4.5$\mu$m, respectively) \citep{and11}. These observations revealed a significant infrared (IR) excess for SN 2010jl, suggesting that a large amount of dust may exist around this supernova before explosion.

In this paper, we present extensive observations of the SN IIn 2010jl in optical bands, providing another well-observed example with which to compare other SNe IIn. \citet{smith12} and \citet{sto11} have studied some of the optical properties of SN 2010jl, but our excellent and independent dataset of both photometry and spectroscopy allow us to provide better constraints on the properties of SN 2010jl. Our observations and data reduction are described in $\S$ 2, while $\S$ 3 presents the light curves and spectra, and the spectral evolution (especially the hydrogen emission lines) is given in $\S$ 4. In \S 5 we present the discussions about the progenitor of SN 2010jl. Our summaries and conclusions are given in $\S$ 5.

\section{Observations and data reduction}

\citet{np10} discovered SN 2010jl with an unfiltered magnitude of $\sim$ 13.0 mag on 2010 November 02.06 UT at $\alpha$ = 09$^{\rm{h}}$42$^{\rm{m}}$53$^{\rm{s}}$.33, $\delta$ = +09$^{\circ}$29'41''.8 (J2000.0). It exploded at 2".4 east and 7".7 north of the center of the nearby irregular galaxy UGC 5189A (see Figure 1). The host galaxy appears very diffuse, consisting primarily of low-metallicity stellar populations (e.g., \citet{sto11}). An optical spectrum taken on Nov. 5.08 showed that SN 2010jl is consistent with a type IIn supernova at the early phase, with prominent H$\alpha$ emission lines and visible helium lines \citep{ben10}. Owing to the brightness and rarity of SNe IIn events, we requested frequent optical ($UBVRI$ bands) imaging as well as optical spectroscopy and collected a total of $\sim$ 300 photometric data points and 12 optical spectra until May 2012.

\subsection{Photometry}

The photometry of SN 2010jl was obtained by the 0.8-m Tsinghua-NAOC reflecting Telescope (TNT\footnote{This telescope is co-operated by Tsinghua University and the National Astronomical Observatories of China (NAOC)}) located at NAOC Xinglong Observatory. This telescope is equipped with a 1340 $\times$ 1300 pixel back-illuminated CCD, with a field of view (FOV) of 11.5' $\times$ 11.2' (pixel size $\sim$ 0.52'' pixel$^{-1}$). Our observation of SN 2010jl started on 2010 November 8, about 6 days after the discovery, and extended out to about 500 days after that.

As shown in Figure 1, SN 2010jl is located close to the center of UGC 5189A. Since this supernova fades slowly and is still quite bright ($\sim$ 16.0 mag) in 2012, we did not apply an image-subtraction technique before performing photometry. The aperture photometry was performed to obtain the instrumental magnitudes of both the SN and the local standard stars with the pipeline for the TNT data reduction (based on the IRAF\footnote{IRAF, the Image Reduction and Analysis Facility, is distributed by the National Optical Astronomy Observatories, which are operated by the Association of Universities for Research in Astronomy (AURA), Inc., under cooperative agreement with the National Science Foundation.} DAOPHOT package). To estimate the contamination of the background light from the host galaxy, we used the pre-explosion image taken by the Sloan Digital Sky Survey (SDSS; e.g. \citet{am08}). The fiber magnitude measures the total flux contained within the aperture of a spectroscopic fiber (3" in diameter), and is calculated in BVRI band at the position of SN 2010jl. We found that the background emission from the host galaxy is fainter than 18.0 mag in each band. It is therefore assumed that the light from the host galaxy barely affects the early-time photometry of SN 2010jl, but it could cause an uncertainty up to about 5\% in the photometry when the SN becomes fainter than 15.0 mag.

A series of standard stars \citep{land92} covering a wide range of colors and air masses were observed on photometric nights to transform the instrumental magnitudes to the standard Johnson $UBV$ \citep{john66} and Kron-Cousins $RI$ \citep{cou81} systems. A total of four photometric nights were used to calibrate the local standard stars in the field of SN 2010jl (see finding chart in Fig. 1). The average values of the color terms obtained on different photometric nights are consistent with previous estimates for the TNT photometric system (e.g., \citet{wxf08, wxf09, ztm10, huang12}), which are not listed here. The final calibrated $UBVRI$ magnitudes of eight standard stars are listed in Table 1. These stars are then used to transform the instrumental magnitudes of SN 2010jl to the standard $UBVRI$ magnitudes, with the final results of the photometry listed in Table 2.
The error bars (in parenthesis) include both the uncertainty in the calibration of the local standard stars and the uncertainty of the instrumental magnitudes.

\subsection{Spectroscopy}

Low-resolution optical spectra of SN 2010jl were obtained with the Cassegrain spectrograph and BAO Faint Object Spectrograph \& Camera (BFOSC) mounted on the 2.16-m telescope at NAOC Xinglong Observatory. Two late-time spectra taken at 86 days and 392 days after maximum light were, respectively, obtained with the Carelec spectrograph on the 1.93-m telescope at Haute-Provence Observatory (OHP) France and the Yunnan Astronomical Observatory Faint Object Spectrograph \& Camera (YFOSC) on the 2.4-m telescope at Lijiang Observatory. A journal of spectroscopic observations is listed in Table 3.

All spectra were reduced by IRAF routines. In order to avoid contamination from the host galaxy, each spectrum of the SN was extracted carefully, and was calibrated with the spectrophotometric standard stars observed on the same night at similar air masses as the SN. The spectra were corrected for the atmospheric extinction using mean extinction curves for the local observatories; moreover, telluric lines were also removed from the data.

\section{Light curves of SN 2010jl}

\subsection{Optical Light Curves}

The $UBVRI$ optical light curves of SN 2010jl are shown in Figure 2. Overplotted are the $V-$ and $I-$band light-curve data from the All-Sky Automated Survey (ASAS) North and Swope \citep{sto11} which provides early-time observations starting from about 25 days before the discovery. These complementary data at early times allow us to yield $V_{max}$ = 13.8$\pm$0.1 mag on JD 2455488.0. This indicates that our observations started from t $\sim$ 20 days and extended to t $\sim$ 500 days, with respect to the $V$-band maximum (hereafter the maximum is referred to the $V$-band maximum). The $I$-band light curve reached a peak magnitude of 13.0$\pm$0.1 mag on JD 2455494, about one week after the t$_{V_{max}}$. Adopting a distance modulus $\mu$ = 33.43$\pm$0.15 mag from NED\footnote{NASA/IPAC Extragalactic Database: http://nedwww.ipac.caltech.edu/} and a galactic reddening $E(B - V)$ = 0.027 mag \citep{sch98}, the peak absolute magnitudes are derived as M$_{V}$ = $-$19.9 mag, M$_{I}$ = $-$20.5 mag, and M$_{R}$ $<$ $-$20.0 mag, respectively. It is clear that SN 2010jl is much brighter than a normal SN IIn which has a typical peak magnitude in the range of M$_{R}$ =$-$17.0 to $-$18.5 mag according to a recent  statistical study by \citet{kie12}, which puts it into the luminous group of type IIn supernovae (see also Figure 4).

The overall light-curve evolution of SN 2010jl is quite similar in the $UBVRI$ bands, showing an interesting two-stage, linear decline after the peak. Before t $\sim$ 90 days from the maximum, the decay rates measured per 100 days in each band are 1.17$\pm$0.03 mag in $U$, 0.96$\pm$0.02 mag in $B$, 0.92$\pm$0.01 mag in $V$, 0.74$\pm$0.01 mag in $R$, and 0.88$\pm$0.01 mag in $I$, respectively. This post-maximum decay rate seems to be wavelength dependent, with faster declines in bluer bands. An exception occurs in the $R$ band where the decay rate is unusually small (see Figure 4 for a comparison with other SNe IIn), which may be due to the strong interactions with the dense hydrogen-rich CSM began shortly after the explosion of SN 2010jl. This mechanism may also explain for the late-time evolution of the light curves in other bands. One can see from Figure 2 that the $UBVRI$ light curves flatten out after t $\sim$ 90 days since the maximum, with the decline rates measured per 100 days as 0.22$\pm$0.02 mag in $U$, 0.22$\pm$0.03 mag in $B$, 0.27$\pm$0.02 mag in $V$, 0.13$\pm$0.01 mag in $R$, and 0.32$\pm$0.01 mag in $I$, respectively. These decay rates are much smaller than the radioactive decay rate of the $^{56}$Co $\rightarrow$ $^{56}$Fe [e.g., 0.98 mag (100 days)$^{-1}$]. This requires an input of energy from other sources such as the interaction of the SN ejecta with the CSM, which is in accord with the prominent hydrogen emission observed in the optical spectra (see Figure 6).

\subsection{Color Curves}

Figure 3 shows the corresponding color evolution of SN 2010jl. Overplotted are some well-observed, luminous SNe IIn sample, including SN 1997cy \citep{ger00}, SN 1998S \citep{fas00}, SN 2006tf \citep{smith08}, and SN 2006gy \citep{smith07a}. In $U-B$, SN 2010jl evolved redwards very slowly, e.g., from $-$0.5 mag at t $\sim$ 20 days to $-$0.4 mag at t $\sim$ 200 days. The $U - B$ color evolution is not known for other comparison SNe for the lack of the photometric data in the $U$ band. In $B - V$ and $V - I$, the color curves kept nearly a constant for SN 2010jl during the phase from t $\sim$ 20 days to t $\sim$ 200 days after the maximum. This suggests that the shape of the continuum spectra did not change significantly during this period, which is consistent with that the blackbody temperature inferred from the spectra stays at $\sim$ 7000 K at similar phases (see Table 5). Inspecting Figure 3 reveals that SN 2006tf might have similar colors as SN 2010jl at the early times but it gets gradually redder at a pace of 0.25$\pm$0.03 mag (100 day$^{-1}$) in $B - V$. While SN 1998S has much bluer $B - V$ and $V - I$ colors in the early phase but it evolved rapidly towards red colors until t $\sim$ 70 days after the maximum light. The $V - R$ color, which can be used to roughly measure the relative strength of the H$\alpha$, also exhibits remarkable differences for the four sample in our study. SN 2010jl and SN 1998S became progressively red after maximum light, with a slope of 0.16$\pm$0.01, and 0.83$\pm$0.02 (100 days)$^{-1}$, respectively, while SN 2006tf and SN 1997cy remained almost a constant $V - R$ color at similar phases. This discrepancy indicates that an intense interaction with the CSM may exist in the former two objects while it should be relatively weak for the latter two objects.

\subsection{Absolute Magnitudes and Bolometric Light Curves}

In Figure 4, we compare the absolute $R$-band light curves of SN 2010jl with other SNe samples. The pre-maximum data points are extrapolated through the $V$-band light curves published by \citet{sto11} and the linear $V - R$ color curve as presented in Figure 3. Overplotted are the light curves of a typical type Ia SN 2005cf \citep{wxf09} and type IIP SN 1999em \citep{leo02}. The hubble constant $H_{0}$ is taken to be 72 km s$^{-1}$ Mpc$^{-1}$ \citep{free01} in deriving the absolute magnitudes for these objects. Days of the light curves are given relative to the estimated phase of the maximum brightness. One notable feature of this plot is the heterogeneous light curves of the type IIn SNe. Compared with type Ia and IIP SNe, the SNe IIn samples are very luminous and their high luminosity lasts for a long time after the peak. A recent study by \citet{kie12} suggests that the rise time for those luminous core-collapse can be very long ($>$ 20 days). As seen from the plot, SN 2006gy could have a rise time longer than 60 days, which is the most luminous SNe IIn ever recorded in recent years. Although less extreme relative to SN 2006gy, SN 2010jl clearly shows a high luminosity with a long duration. It is interesting to note that the light curve of SN 2010jl is similar to that of SN 2006tf and SN 1997cy at early times, but remarkable difference between them emerges at late times. By t $\sim$ 90 days from the maximum light, the light curve of SN 2010jl becomes almost flat, unlike SN 1998S or SN 2006tf which declined in $R$ at a faster pace at similar phase. By t $\geq$ 200 days, SN 2010jl becomes the most luminous SN IIn of our samples due to its remarkably slow evolution at late times. Such a long-duration emission at high luminosity demands of a large amount of emitting materials. Possible energy sources include $^{56}$Co decay from a large amount of $^{56}$Ni, diffusion of the shock-deposited kinetic energy, and CSM interactions (see the discussions of other possible energy sources and their difficulties in \citet{smith07a}).

To better understand the overall properties, and particularly the energy sources of SN 2010jl, we constructed its quasi-bolometric light curve using the $UBVRI$ photometry presented in this paper. In this calculation, we used the normalized passband transmission curves given by \citet{bes90}. The integrated flux in each filter was approximated by the mean flux multiplied by the effective width of the passband, and was corrected for the reddening in the Milky Way. As we did not get any ultraviolet (UV) or near-infrared (NIR) data for SN 2010jl, we estimate their contribution in the following ways. From the spectral energy distribution of SN 2010jl constructed at t $\sim$ 90 days after maximum by \citet{and11}, we estimate the contribution from the infrared (IR) emission to be less than 15\% of the total flux. Moreover, \citet{and11} suggested that the IR emission should come primarily from the radiation of the pre-existing dust heated by the SN photons. Thus the actual contribution of the IR emission to the total flux of the SN could be much lower than the number quoted above. Assuming that the continuum of SN 2010jl can be well represented by a blackbody spectrum, e.g., with an effective temperature of about 7,000 K, the UV contribution is estimated to be less than 10\% of the total flux.

Figure 5 shows the bolometric light curve of SN 2010jl, derived from the $UBVRI$ light curves. The dashed line shows the $^{56}$Co luminosity and the decay rate expected from an initial nickel mass of about 3.4 M$_{\odot}$. This nickel mass is slightly larger than that estimated for SN 1997cy (e.g., $\sim$ 2.3 M$_{\odot}$; \citet{tur00}) but smaller than that of SN 2006tf (e.g., $\sim$ 4.5 M$_{\odot}$; \citet{smith08}). One can see that the radioactive decay of the cobalt gives a better fit to the bolometric light curve before t $\sim$ 90 days from the maximum, but it deviates from the observed light curve after that. The late time luminosity of SN 2010jl is much higher than that predicted by the radioactive decay of $^{56}$Co, which requires an input of additional energy such as the CSM interaction. In the study of SN 2006tf, \citet{smith09} suggested that a model with photon diffusion from a massive shocked envelope may provide a better match to the main light curve peak like that proposed for that of SN 2006gy \citep{smith07b}. In such a model, the extreme luminosity came from the thermalization of the CSM envelope by the shock kinematic energy. The dash-dotted line indicates the luminosity evolution for a diffusion model with a diffusion timescale t$_{diff}$ = 160 days from \citet{smith08}. It is clear that this model does not match to the observed data of SN 2010jl at later phase, which predicts a late-time decline rate that is even steeper than that of the radioactive decay. We thus propose that a large portion of the late-time emission of SN 2010jl may be supplied by the CSM interaction as also indicated by the astonishing growth of the H$\alpha$ and H$\beta$ emission in the late time spectra (see discussions below). The dotted line in Figure 5 represents the luminosity produced from a simple SN ejecta-CSM interaction model with L $\propto$ t$^{-3\alpha}$ \citep{wood04}, where $\alpha$ is a parameter to describe the power-law ejecta density \citep{cf94}. The best fit to the late time curve yields $\alpha$ = 0.1, which is well within the typical range of a power-law distribution for normal type II SNe as given by \citet{cf94}.

We thus propose that both the radioactive decay of a large amount of $^{56}$Co and the extreme CSM interaction may contribute to the high luminosity of SN 2010jl. Compared with SN 2006tf, SN 2010jl may have a slightly faint bolometric luminosity near maximum light but it probably suffered a more extreme CSM interaction which leads to a higher luminosity at late times. We also integrated this quasi-bolometric light curve over the period spanning approximately from the beginning of explosion to t $\sim$ 200 days after the maximum light. The total energy radiated in the visual light during this period is estimated to be 4.3$\pm$0.2$\times$10$^{50}$ ergs, which is comparable to SN 2006tf (6.2$\times$10$^{50}$ ergs) considering the uncertainties in the calculations.

\section{Evolution of Optical Spectra}

A total of twelve optical spectra of SN 2010jl were obtained at NAOC Xinglong Observatory, YNAO Lijiang Observatory and Haute-Provence Observatory, spanning t = 24 to 515 days relative to the $V$-band maximum date (JD = 2,455,488; \citet{sto11}). Figure 6 presents the complete spectral evolution. Following \citet{smith08}, we normalized our spectra to the red continuum flux and offset them by constant values for a better display. Given that Na ID lines are very weak in the spectra, we only applied an absorption correction to our spectra with the Galactic reddening of $E(B - V)$ = 0.027 mag. The most remarkable feature about the spectral evolution of SN 2010jl is the astonishing growth of the equivalent width of H$\alpha$ and H$\beta$ with time, and is to be discussed in detail in \S 4.2. Note that the emission features near 5000 {\AA} are due to the forbidden lines [OIII] $\lambda\lambda$4960, 5010{\AA} from the host galaxy near the SN position, as shown at the bottom of figure 6. We discuss in detail the spectroscopic evolution of SN 2010jl in the following subsections.

\subsection{Temporal Evolution of the Spectra}

A spectroscopic comparison with some notable type IIn supernovae such as SNe 1997cy \citep{tur00}, 1998S \citep{fas01, poz04}, 2006gy \citep{smith10}, and 2006tf \citep{smith08}, at several epochs (t $\approx$ 1 month, 3 months, 6 months, and 15 months past the maximum) is shown in Figure 7. All the spectra have been corrected for the Galactic reddening and redshift of the host galaxy.

The comparison of the spectra at t $\approx$ 1 month is shown in Figure 7a. Pronounced emission features of balmer lines already exist in the earliest spectrum of SN 2010jl, which appear to consist of a narrow component and an intermediate-width component. Besides the Balmer lines, the He I emission lines at 5876 and 7065 {\AA} are also visible in the spectra and a narrow component also exists in these lines. Of the four comparison SNe IIn, SN 2010jl shares most of the spectral features with SN 2006tf except that it has a more pronounced helium lines. SN 2006gy displayed a spectrum similar to SN 2010jl and SN 2006tf, but with relatively weak features of H and He. SN 1998S seems to have a different spectral characteristic at this phase, with much stronger, bluer continuum and shallower absorption profiles, reminiscent of those of a normal type II \citep{fas00}. While SN 1997cy is an extraordinary SN IIn probably associated with the gamma-ray burst. Its t $\sim$ 1 month spectrum is characteristic of lines of different widths, especially the broad absorption lines in the 4500-6000 {\AA} region, which is formed perhaps due to blends of several lines at higher velocities.

The comparison at t $\approx$ 3 months is shown in Figure 7b. The continuum shape and overall properties of the spectrum do not show significant change since t $\sim$ 1 month, except for SN 1998S (which appears much redder relative to the earlier spectrum). One significant change in the t $\sim$ 3 months spectrum is the remarkable growth of the the H$\alpha$ feature with time. For example, the FWHM velocity of the intermediate-width component changes from $\sim$ 2000 km s$^{-1}$ to 3000 km s$^{-1}$ for SN 2010jl. Such a strengthening of H$\alpha$ emission have been observed in other comparison SNe IIn such as SN 1998S which showed the most striking contrast of the flux ratio between t $\sim$ 1 month and $\sim$ 3 months. The most interesting change in the plot is that an asymmetric line profile develops in the intermediate-width component of H$\alpha$, with the blue side being much stronger than the red side (see also Figure 8 and discussions in \S 4.2). Such an asymmetric line profile was also seen in the spectra of all of the comparison sample at similar phases. This has been explained by post-shock dust formation inside the SN ejecta, which may block more emission from the further side of the SN (e.g., \citet{smith09}). However, this explanation may be in contradiction with the evolution of late time light curve of some SNe IIn, at least for SN 2010jl (see discussions below). At t $\approx$ 3 months, we also notice that the emission of Ca II IR triplet became strong in SN 1997cy and SN 1998S, but it was very weak in SN 2006tf. This indicates that SN 1997cy and SN 1998S may have faster expanding ejecta and their inner parts are exposed earlier to the observers if compared to SN 2006tf. We can not judge the case for SN 2006gy and SN 2010jl from our spectrum because of a shorter wavelength coverage. However, a spectrum taken at a similar epoch from \citet{smith12} shows that a relatively weak feature due to IR Ca II triplet can be seen for SN 2010jl.

In Figure 7c, we compare the spectrum of SN 2010jl with those of SN 1997cy and SN 2006gy at about half a year from the maximum. By t $\sim$ 6 months, the Balmer lines of SN 2010jl continued to gain strength and now are much stronger than those of the comparison objects, while the He lines remained almost unchanged. For SN 1997cy and SN 2006gy, both the continuum and the relative strength of H$\alpha$ do not show significant change during this period (see also Figure 9). The Ca II IR feature gets much stronger in SN 2010jl, comparable to that of SN 1997cy. While this feature is very weak and almost invisible in the t $\sim$ 6 months spectrum of SN 2006gy.

In Figure 7d, we compare the spectra at t $\approx$ 15 months. The spectrum at this phase is dominated by the outstanding emission lines of H, He, and Ca II IR triplet, indicative of a successive, strong CSM interactions. SN 1998S is the only comparison SN IIn with such a late time spectrum, which also displayed a very strong feature of H$\alpha$ emission with multiple components. We note that a prominent absorption feature appears at $\sim$ 5000 {\AA} in the spectrum of SN 2010jl, which is likely due to Fe II $\lambda$5169. The appearance of noticeable iron lines in SN 2010jl is relatively late with respect to all of the four comparison SNe IIn. This indicates that SN 2010jl may have thicker outer shells and it takes longer time to expose the innermost iron core to the observers. The presence of protruding forbidden lines [OIII] $\lambda\lambda$ 4960, 5010 {\AA} in the SN spectrum reminds us, however, that the late time spectrum suffered considerable contamination from the host galaxy.

One can see from the above comparison that the most impressive spectroscopic characteristics for SN 2010jl are the tremendous growth of the strength of the H lines and the appearance of asymmetric line profiles at some stage. The quantitative analysis of the evolution of the strength and line profiles are given in the next two subsections. In contrast with the peculiar evolution of the line features, the overall shape of the continuum spectra stays roughly unchanged at different epochs. A fit to the observed spectra yields a blackbody temperature ranging from about 7300 K at t $\sim$ 20 to about 6500 K at t $\sim$ 400-500 days since the maximum. This is consistent with a nearly constant $B - V$ and $V - I$ color evolution inferred from the photometry (see Figure 3). In general, the overall spectroscopic evolution of SN 2010jl resembles closely to that of SN 2006tf \citep{smith08} during the period from t $\sim$ 1 to 3 months.

\subsection{Evolution of H$\alpha$ and H$\beta$ Emission}

Figure 8 shows the evolution of the equivalent width and luminosity of H$\alpha$ and H$\beta$ emission for SN 2010jl. The flux of H$\alpha$ emission is normalized to the nearby continuum, and that of H$\beta$ is scaled for comparison. The observed line profiles of H$\alpha$ emission can be well decomposed into a narrow component with the FWHM velocity $\lesssim$ 1000 km s$^{-1}$, and a broader component with FWHM velocity $\sim$ 2000-3500 km s$^{-1}$. This is achieved by using a double-lorenz/double-gaussian function with separate central wavelengths. Such a two-component fit also applies to the H$\beta$ profiles, as shown in the right panels of Figure 8.

The line profiles of H$\alpha$ and H$\beta$ were symmetric in day 24 and 45 spectra, but they became asymmetric in the spectra after that. The appearance of an enhanced, broader emission on the blue side of the narrow component makes the profiles appear asymmetric. The narrow component of H$\alpha$ shows little changes in both the FWHM velocity and the central wavelength, and is still visible in t $\approx$ 392 days spectrum. The narrow component of H$\beta$ becomes nearly undetectable in the spectra after t $\sim$ 200 days. It is generally believed that this unshifted narrow component was produced by circumstellar gas lost from the progenitor prior to explosion and then photo-ionized by the UV radiation emitted at the time of shock breakout. At late times, this circumstellar gas should be eventually engulfed by the expanding SN ejecta. In contrast with the narrow component, the intermediate-width component of Balmer lines exhibit a complex evolution. The center of H$\alpha$ is systematically blueshifted, with the doppler velocity ranging from $\sim$ $-$300 km s$^{-1}$ on days 24 to $\sim$ $-$800 km s$^{-1}$ on days 515. The H$\beta$ line seems to experience similar evolution, with the corresponding velocity changing from $\sim$ $-$100 to $\sim$ $-$700 km s$^{-1}$. Moreover, the FWHM velocity of the intermediate-width components varied between 2000 $-$ 3500 km s$^{-1}$. This component can be explained with an CSM shell swept by the forward shock (e.g., \citep{smith08}). However, the fast expanding SN ejecta can catch up the post-shocked CSM shell at some point, which may further heat and accelerate the CSM shell through an interaction. This configuration may account for the continual growth of the equivalent width (EW) and the systematic blueshift of the H$\alpha$ profile as shown in Figure 8.

\citet{smith12} proposed that such an asymmetry of the line profile was likely due to the post-shock dust formation, and they estimated that the new dust grains formed around t $\sim$ 90 days from the maximum. In principle, the formation of new dust inside the ejecta would leave three observable signatures: 1) a remarkable decrease in the decline rate of the optical light curves, especially at shorter wavelengths; 2) systematic blueshift of emission-line profiles when the photons from the receding side are blocked by new dust; 3) infrared excess due to thermal emission from warm or hot dust. Note, however, that infrared excess may not be a reliable tracer of new dust formation since the IR excess could be also caused by the pre-existing dust and radiation efficiency of the new dust is still uncertain. For example, all of the above three signatures were clearly seen in the peculiar type Ib SN 2006jc which was believed to suffer an early dust formation in the ejecta (e.g., \citet{anu09}). Although the asymmetric line profiles are seen in the spectra of SN 2010jl, its $UBVRI$ light curves did not show a sudden decline at similar phases. During the period from t $\sim$ 1 month to t $\sim$ 6 months, the $U - B$ and $V - I$ color also stays roughly a constant (see Figure 3), with no sign of suffering from significant reddening from the new dust. Note that the H$\alpha$ line profile showed an progressive blueshift until t $\sim$ 431 days, but the blueshift seems to decrease at t $\sim$ 515 day and the corresponding line profile became more symmetric at this time. These are at odds with the post-shock formation of new dust. Finally, \citet{and11} provided evidence that the significant IR excess they observed in SN 2010jl is due to an IR echo caused by the pre-existing dust around SN 2010jl, not due to the newly formed dust.

To elucidate these inconsistencies, we propose that such a line asymmetry may be attributed to interactions of post-shock CSM shells with the expanding ejecta. The pre-shocked CSM shell locates far from the supernova, and is responsible for the unshifted narrow component of the line profile. While the inner or post-shock CSM shells, accounting for the intermediate-width component of the line profile, can be successively accelerated and heated during the interaction stage. The superposition of these two components can thus result in an asymmetric profile of the Balmer lines seen in the spectra. At late times, the acceleration for the CSM shells becomes less effective with the consuming of kinematic energy of the ejecta. The intermediate-width component gets less blueshifted and the line profiles become symmetric again, as seen in t$\sim515$ day spectrum of SN 2010jl. Interestingly, we note that the time when the H line starts to become asymmetric is roughly coincidence with the time when the the light curves become flat. This suggests that the interaction should play an important role in producing the asymmetric blueshifted profiles. \citet{smith12} also propose an alternative explanation for such a line asymmetry in light of the high continuum optical depths. With the receding of the photosphere into the post-shock CSM shell, the H emission from the near side of the shock could still reach us but the emission from the far side of the post-shock shell may be blocked by the continuum photosphere. This could also result in the appearance of an asymmetric, blueshifted line profile. Such a hypothesis can be examined by the late-time photometry (e.g., t $\gtrsim$ 500 days), as the transition of the optical depths usually affect the decline rate of the light curve at late times.

To perform a better comparison with the well-known SNe IIn, we calculated some spectroscopic parameters for H$\alpha$ and H$\beta$ and derived their evolution with time, including the equivalent widths (EWs), luminosity, FWHM velocity, blueshift, and flux ratio of these two lines. These results are shown in Figure 9 and also listed in Table 4. As we mentioned before, the contamination of the host galaxy can't be well removed, especially for the narrow component in the emission lines. Therefore, here we focus on the study of the intermediate-width component of the line profiles and intensity. As shown in the upper panel of Figure 9, the EWs of different SNe IIn roughly increase in a linear fashion with the explosion time. Note that a dramatic increase seen in SN 1998S at t $\sim$ 90 days perhaps relates to the strong, broad component from the SN ejecta. Although the H$\alpha$ EW of SN 2010jl is at an intermediate place among the comparison sample at the beginning, it increases continuously at a faster pace with time. At t $\sim$ 515 days, it superseded all the comparison SNe IIn and became the one with the largest EW. The EWs of H$\beta$ are also presented for SN 2010jl and SN 2006tf. SN 2010jl shows a progressive increase in H$\beta$ until t $\sim$ 400 days after maximum, while SN 2006tf shows an increase before t $\sim$ 90 days but stays almost a constant after that. We compare the line intensity in the lower panel of Figure 9. SN 2010jl clearly displays a noticeably high H$\alpha$ luminosity relative to other comparisons, especially at late phase. At t $\sim$ 400 days, its H$\alpha$ luminosity is about 1.0$\times$10$^{42}$ erg s$^{-1}$, which is brighter than those of SNe 1988Z, 1997cy, and 2006tf by about an order of magnitude. The H$\alpha$ flux of SN 2010jl seems to suffer a sudden drop after t $\sim$ 400 days, like that of SN 1997cy, perhaps indicating that the interactions of the SN ejecta with the CSM shells weakened significantly at this stage. This is fully consistent with the discovery of a reduced blueshift of the H$\alpha$ line profile in t $\approx$ 515 day spectrum as discussed above.

We also estimated the flux ratio between H$\alpha$ and H$\beta$. As shown in Table 5, the H$\alpha$/H$\beta$ flux ratio basically fluctuates around a value of $\sim$ 4 at different epochs. This result is not so different from SN 2006tf at early times, but significant difference emerges after 400 days from the maximum. At t $\sim$ 400-500 days, this ratio does not climb to a value $>$ 10 as seen in SN 2006tf at similar phase. \citet{smith08} proposed that a large H$\alpha$/H$\beta$ flux ratio might relate to a transition of the energy mechanism from the recombination emission (photo-ionization heating) to the purely collision excitation. It is thus interesting to explore the reason that SN 2010jl kept a relative strong H$\beta$ emission at such a late time. On the other hand, we must keep in mind that the measurement of H$\alpha$/H$\beta$ ratio as listed in Table 5 may suffer a considerable large uncertainty due to host-galaxy contamination, especially when the emission from the SN fades significantly at late times.

\subsection{He I lines}

Comparing with the prominent Balmer lines, the He I emission lines of SN 2010jl are clearly detected at 5876 and 7065 {\AA} in all of our spectra. The He I feature at 6678 {\AA} is barely visible due to blends with the broad H$\alpha$ lines. With respect to the comparison SNe IIn as shown in Figure 7, SN 2010jl showed relatively stronger He I lines. For the emission feature at 5876-{\AA}, an intermediate-width component with FWHM velocity $\sim$ 2750 km s$^{-1}$ can be identified like that for the Balmer lines. As shown in the lower panel of Figure 9, the measured flux of He I 5876 stayed approximately a constant of about 6.0$\times$ 10$^{40}$ erg s$^{-1}$ until t $\sim$ 200 days from maximum, and it then dropped dramatically after that. Such an evolution trend is similar to that observed for H$\alpha$ at similar phases. This means that He has a uniform distribution in the CSM shell, similar to that of H.

\section{Physical Properties and Progenitor of SN 2010jl}

In the above two sections, we have described the overall evolution of the $UBVRI$ light curves and the optical spectra. The extensive photometric and spectroscopic data allows us to derive the bolometric luminosity, the blackbody temperature as well as the intensity and shift for some characteristic lines such as H$\alpha$ and H$\beta$, which are listed in Table 5. To get more clues to the nature of the progenitor of SN 2010jl, we
further derived the progenitor mass-loss rate and wind speed of CSM shell in the following analysis. Based on a CSM-interaction model from \citet{cf94} and an assumption that the luminosity of the ejecta-CSM interaction is fed by the energy imparted at the shock front, the progenitor mass-loss rate $\dot{M}$ can be calculated using the following expression (e.g., \citet{wood04}):
\begin{equation}
    \dot{M} = 2L\frac{v_{\rm{w}}}{\alpha \times v_{{\rm{s}}}^{3}},
\end{equation}
where $L$ is the observed bolometric luminosity, and $\alpha$ represents the conversion term from kinetic energy to optical luminosity. $v_{\rm{w}}$ is the velocity of the pre-explosion stellar wind, which is adopted as $-$28 km s$^{-1}$ inferred from an absorption minimum of the narrow component of H$\alpha$ \citep{smith11b}. $v_{\rm{s}}$ is the velocity of the post-shock shell, assuming as the FWHM velocity of the intermediate-width component of H$\alpha$ lines (see Table 5). To estimate the mass-loss rate, we need to know the parameter $\alpha$. Recalled in \S 3.3, we applied a fit to the bolometric luminosity with a time dependence luminosity relation $L\propto t^{-3\alpha}$, which gives $\alpha\approx$ 0.1. This implies a conversion coefficient from kinematic energy to luminosity of about 10\%.

Inserting the relevant parameters into equation (1), we can obtain the mass-loss rate $\dot{M}$ as listed in column (4) of Table 5. With the mass-loss rate, we can further estimate the timescale of the stellar wind flow, t$_{\dot{M}}$ = $R_{\rm{s}}/v_{\rm{w}}$, which is the number of years prior to explosion when the progenitor had such a mass-loss rate. $R_{\rm{s}}$ is the radius of the CSM shell expanding approximately at an average speed of about 2750 km s$^{-1}$.

One can see from Table 5 that the mass-loss rate derived for SN 2010jl is about 1-2 M$_{\odot}$. As discussed by \citet{smith08}, the mass-loss rate deduced according to eq.(1) may be overestimated because a "pile-up" effect of radiation from earlier CSM interactions would mimic a larger instantaneous luminosity. Taking into account this mimic effect on the bolometric luminosity, an average mass-loss rate of about 0.8 M$_{\odot}$ may be still needed in about 50 years before explosion, indicating that the progenitor of SN 2010jl suffered a substantial mass loss at its final stage. Such a mass-loss rate is about 10$^{4}$ times stronger than the maximum rate that can be supplied by a line-driven wind of a massive star \citep{smith06}. According to the mass-loss rate and the duration listed in Table 5, the progenitor of SN 2010jl might have an initial mass above $\gtrsim$ 30 $-$ 40 M$_{\odot}$ before undergoing a core-collapse explosion. Meanwhile, \citet{smith11b} estimated the progenitor mass from the $HST$ archival images taken roughly 10 years prior to explosion, and they got a lower limit of 30 M$_{\odot}$. This indicates that SN 2010jl can be likely traced to a progenitor of an LBV star, the only star known to be capable of shedding a large amount of mass during some large outbursts like the 19th eruption of $\eta$ Carinae \citep{smith03}.

We point out that the speed of the stellar wind velocity inferred for SN 2010jl (e.g., $-$28 km s$^{-1}$) is unusually slow compared to many other SNe IIn in the literature (e.g., \citet{kie12}) and the LBV eruptions. For example, this wind speed was measured to be $\sim$ 420 km s$^{-1}$ for SN 2005gl \citep{gal09}, $\sim$ 190 km s$^{-1}$ for SN 2006tf \citep{smith08}, and $\sim$ 200 km s$^{-1}$ for SN 2006gy \citep{smith10}, respectively. Such a slow wind speed is actually consistent with that of an extreme red supergiant. This suggests the progenitor star of SN 2010jl underwent a red supergiant stage when the stellar wind materials gradually accumulated to form the dense, dusty CSM as inferred from a significant IR excess \citep{and11}. In theory, it has been shown that a group of LBV stars with relatively lower luminosity are post-red supergiant stars with masses of 30-60 M$_{\odot}$ \citep{stot96}. Thus we propose that SN 2010jl may explode at the post-red supergiant stage of a massive star, which was rarely suggested for SNe IIn before. It is apparent that more sample with accurate analysis of the CSM emission lines are needed to address the fraction of the type IIn explosion like SN 2010jl.

\section{Summary and Conclusion}

This paper presents the photometric and spectroscopic observations of the luminous type IIn Supernova 2010jl. Our data were mainly collected with the 0.8-m TNT and the 2.16-m telescope locating at the Xinglong Observatory of NAOC, spanning a period of about 1.5 years after the explosion. The main results inferred from our data and analysis for SN 2010jl and its progenitors are summarized below.

1. The $UBVRI$ light curves underwent a two-stage evolution, with a break at t $\sim$ 90 days from the maximum. Before t $\sim$ 90 days, the decay rates are in a range of about 0.7 $-$ 1.2 mag (100 days)$^{-1}$, with steeper slope in the bluer bands; while they become much smaller after t $\sim$ 90 days from maximum due to that the light curves flatten out during this period, with values of 0.1-0.3 mag (100 days)$^{-1}$. Such a notable change in the decay rates likely relates to a significant contribution from CSM-interactions at the late phases, as also indicated by the strong emission of H$\alpha$ lines seen in the late-time spectra of SN 2010jl. We also constructed the pseudo-bolometric light curve with our multicolor photometry. The light curve at early phases (t $<$ 90 days) can be well fit by a hypothetical $^{56}$Co decay of 3.4 M$_{\odot}$, while the late-time evolution complies with an ejecta-CSM interaction model. This perhaps indicates that radioactive decay powers SN 2010jl at early phase and CSM interaction provides the energy at later phase.

2. We have also studied some of the important spectroscopic characteristics of SN 2010jl such as H$\alpha$, H$\beta$, and He I lines. These lines, especially the H lines appear very strong in the spectra of SN 2010jl. For example, the H$\alpha$ emission is estimated to have a luminosity of about 0.6-1.0$\times$10$^{42}$ erg s$^{-1}$ before t $\sim$ 400 days. This is apparently stronger than that of other type IIn in the literature, especially at late phase. Another interesting feature about the line profiles of H$\alpha$ and H$\beta$ is that both can be well decomposed into a narrow component and intermediate-width component. The narrow component is likely produced by the outer, pre-shock CSM shell, with an FWHM velocity of about 700 km s$^{-1}$; the intermediate-width component is perhaps due to the post-shock CSM shells, with an FWHM velocity of $\sim$ 2000-3400 km s$^{-1}$, depending on the phases. The intermediate-width component showed a progressively increased strength and blueshift with time until t $\sim$ 400 days. However, an decrease in the strength and the blueshift of the H$\alpha$ and H$\beta$ lines was also noticed in the t $\sim$ 515 day spectra. These changes, together with the flat evolution of the late-time light curves, suggest that the systematic blueshift of the intermediate-width component of the H lines may not be due to the newly formed dust in the post-shock region. Instead, the above changes of the line profiles could be a natural result of CSM-interactions: the intermediate-width component of H lines will appear stronger and more blueshifted with the acceleration of the CSM shells via the successive collision from the ejecta at early times, while the strength and blueshift of this component would decrease when the CSM-interactions became weak at late times. Nevertheless, a detailed quantitative analysis is needed for a test of such a hypothesis.

3. With our extensive light curves and spectral data, we further derived the physical parameters such as the mass-loss rate and stellar-wind flow timescale which allow us to place better constraints on the nature of the progenitor of SN 2010jl. The high mass-loss rate (e.g. $\sim$ 1 M$_{\odot}$) and a short flow timescale of the stellar wind (e.g., $\sim$ 40 years), obtained for the progenitor of SN 2010jl, indicate that the progenitor star must undergo a tremendous amount of mass loss within a few decades before explosion. This makes us believe that SN 2010jl may have a massive progenitor like LBVs. However, the fact that the pre-shock CSM has a slow expansion velocity (e.g., $\sim$ 28 km s$^{-1}$) comparable to that of some red supergiants suggests that SN 2010jl may explode at a post-red supergiant stage. The post-red supergiant can correspond to a group of LBVs with lower luminosity and a mass range of 30-60 M$_{\odot}$. Thus a post-red supergiant progenitor might be possible for SN 2010jl.

\acknowledgments We are grateful to Yang Yanbin for helpful discussions, Lin Zhixing for the assistant of observations, and Feng Qichen for sharing his observation time on the 2.16-m telescope of NAOC. This work has been supported by the Young Researcher Grant of National Astronomical Observatories, Chinese Academy of Sciences. The work of X. Wang is supported by the National Natural Science Foundation of China (NSFC grants 11073013, 11178003), the Foundation of Tsinghua University (2011Z02170), and the Major State Basic Research Development Program (2009CB824800). C. Wu is supported by the NSFC 10903010. Z. Fan is supported by the NSFC 11003021. The work of X. Zhou is supported by the NSFC 11073032.

\clearpage

\begin{figure}[ht]
\centering
\includegraphics[angle=0,width=140mm]{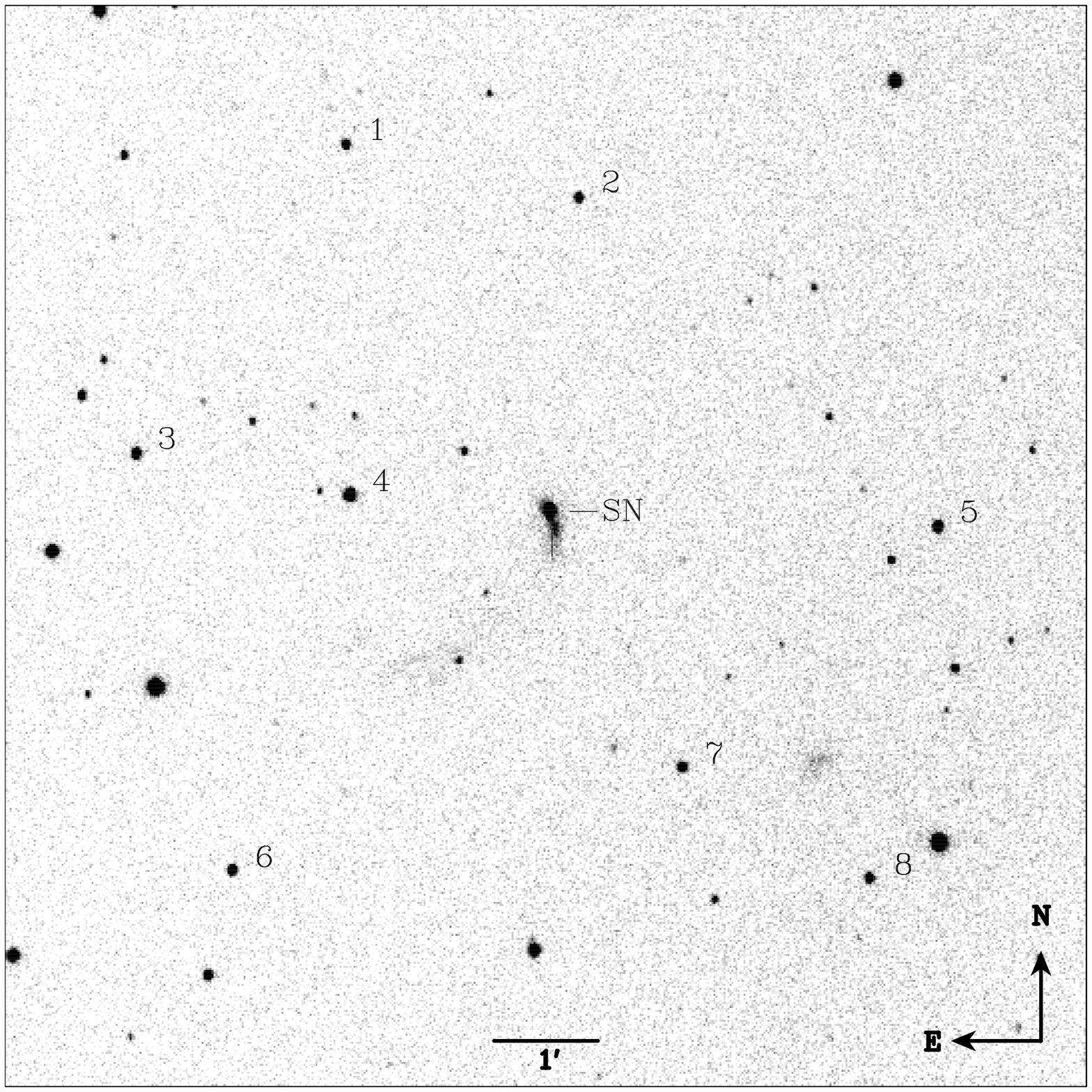}
\caption{SN 2010jl in UGC 5189A. This is a $R$-band image taken with the 0.8-m Tsinghua-NAOC telescope
on 2010 November 16th. The supernova and local reference stars are marked. North is up, and east is to the left.}
\end{figure}

\begin{figure}[ht]
\centering
\includegraphics[angle=0,width=140mm]{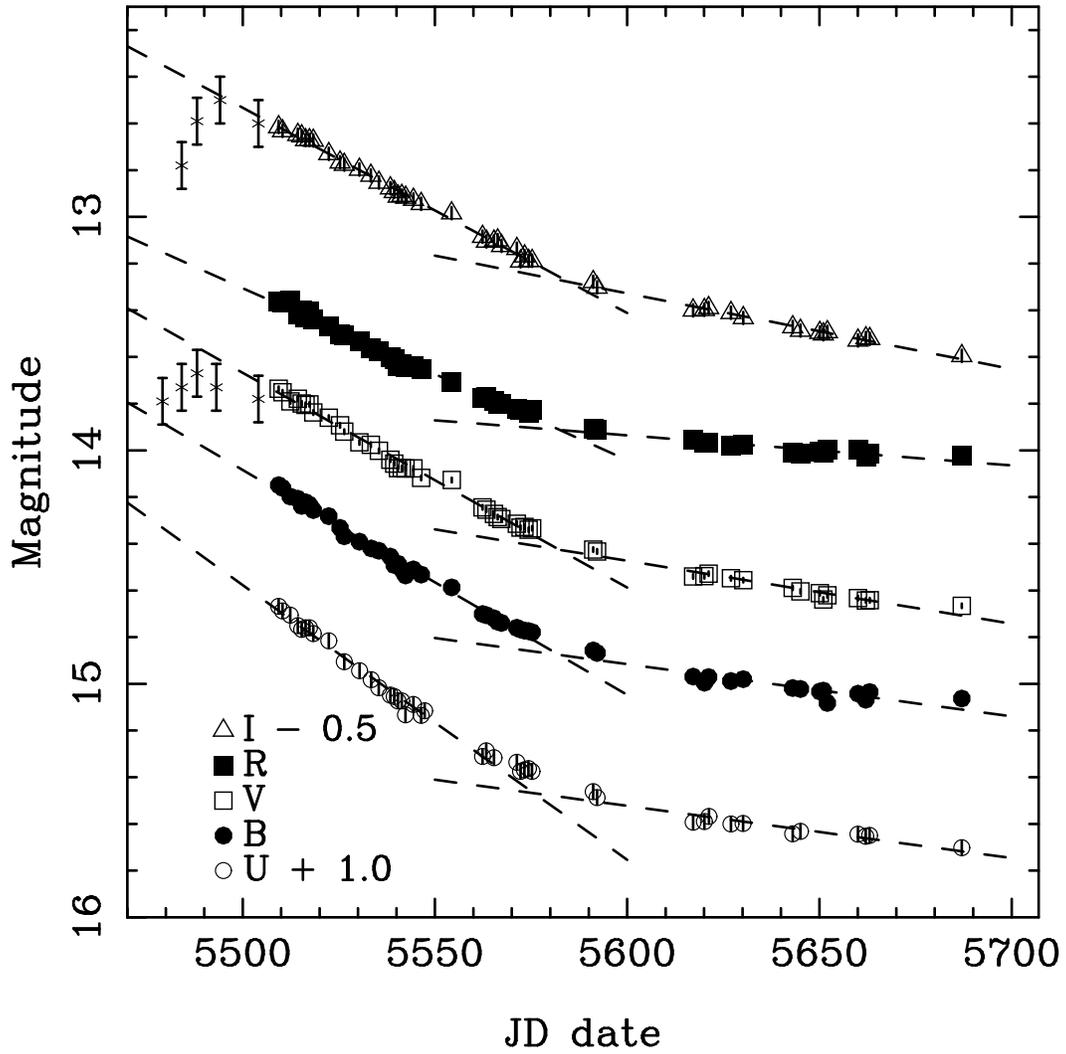}
\caption{The optical $UBVRI$ light curves of SN 2010jl obtained with TNT (see Table 2). The star symbols
are the early $V$- and $I$-band data taken from \citet{sto11}. The dash lines are linear fits to the observed data.}
\end{figure}

\begin{figure}[ht]
\centering
\includegraphics[angle=0,width=140mm]{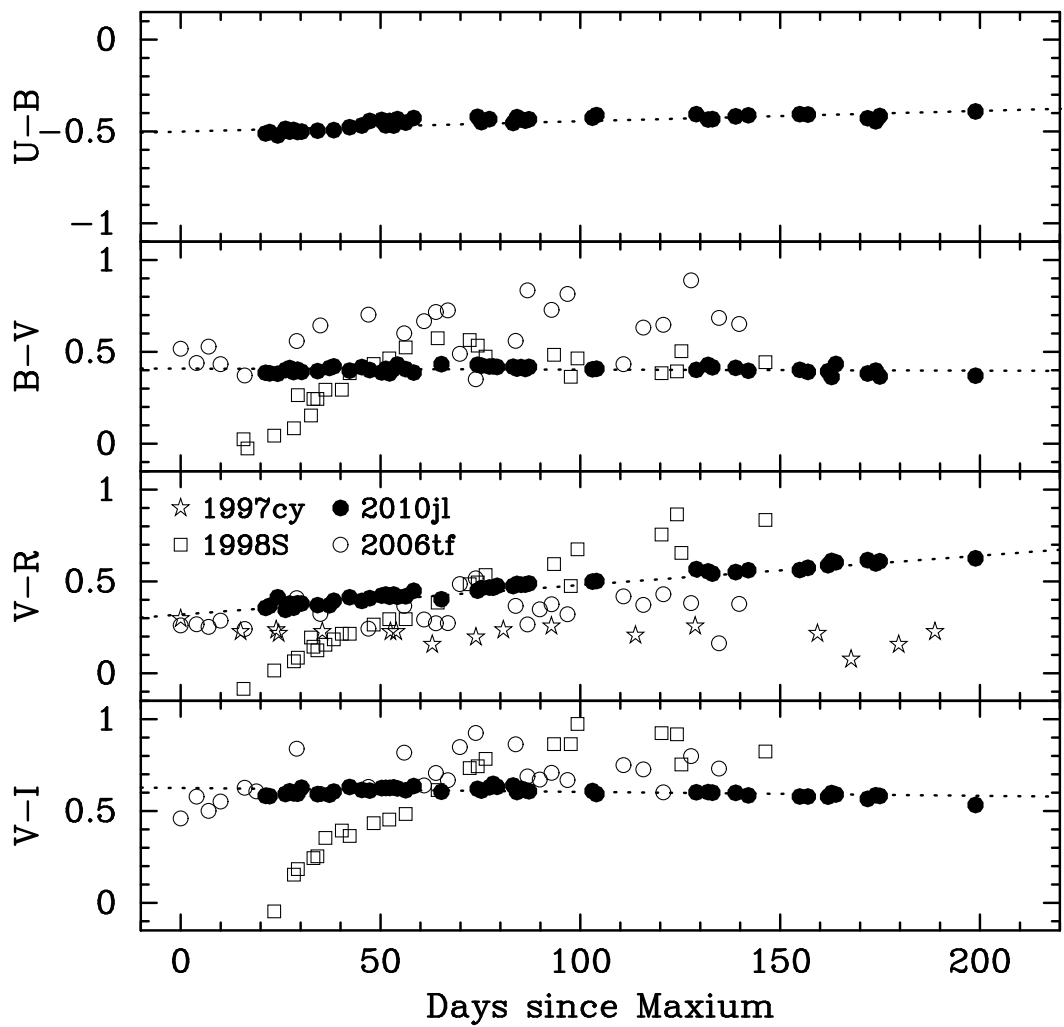}
\caption{$U-B$, $B-V$, $V-R$, and $V-I$ color curves of SN 2010jl compared with those of type IIn
SNe 1997cy \citep{ger00}, 1998S \citep{fas00} and 2006tf \citep{smith08}. All of the comparison SNe have been
dereddened for the reddening in the Milk Way. The dotted lines are least-squares fit to the colors of SN 2010jl.}
\end{figure}

\begin{figure}[ht]
\centering
\includegraphics[angle=0,width=140mm]{f4.ps}
\caption{The absolute $R$-band light curve of SN 2010jl compared with other notable type IIn SNe 1997cy \citep{ger00}, 1998S \citep{fas00}, 2006gy \citep{smith07a} and 2006tf \citep{smith08}, and type Ia SN 2005cf \citep{wxf09}, type IIP SN 1999em \citep{leo02}.}
\end{figure}

\begin{figure}[ht]
\centering
\includegraphics[angle=0,width=140mm]{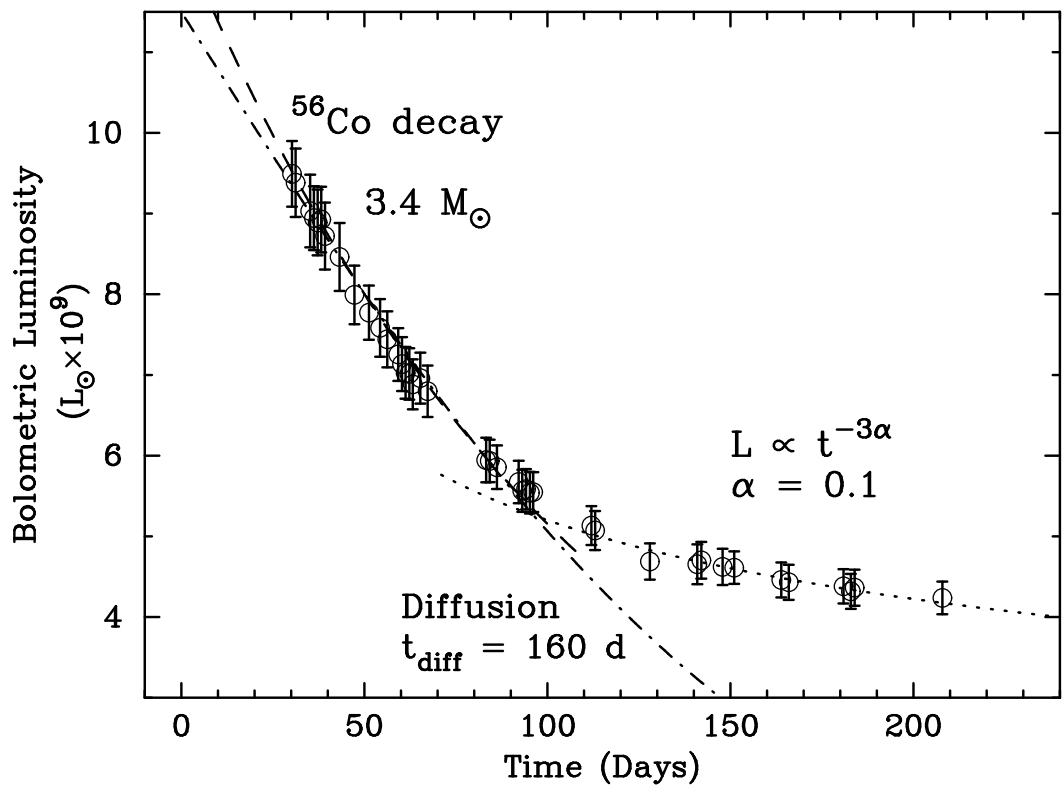}
\caption{The pseudo-bolometric light curve of SN 2010jl. The dashed represents a fit with a radioactive decay of $^{56}$Co of 3.4 M$_{\odot}$. The dot line is the best fit using a simple time dependent luminosity model and dash-dotted line is a simple diffusion model with a timescale t$_{diff}$ = 160 days (see text for details).}
\end{figure}

\begin{figure}[ht]
\centering
\includegraphics[angle=0,width=140mm]{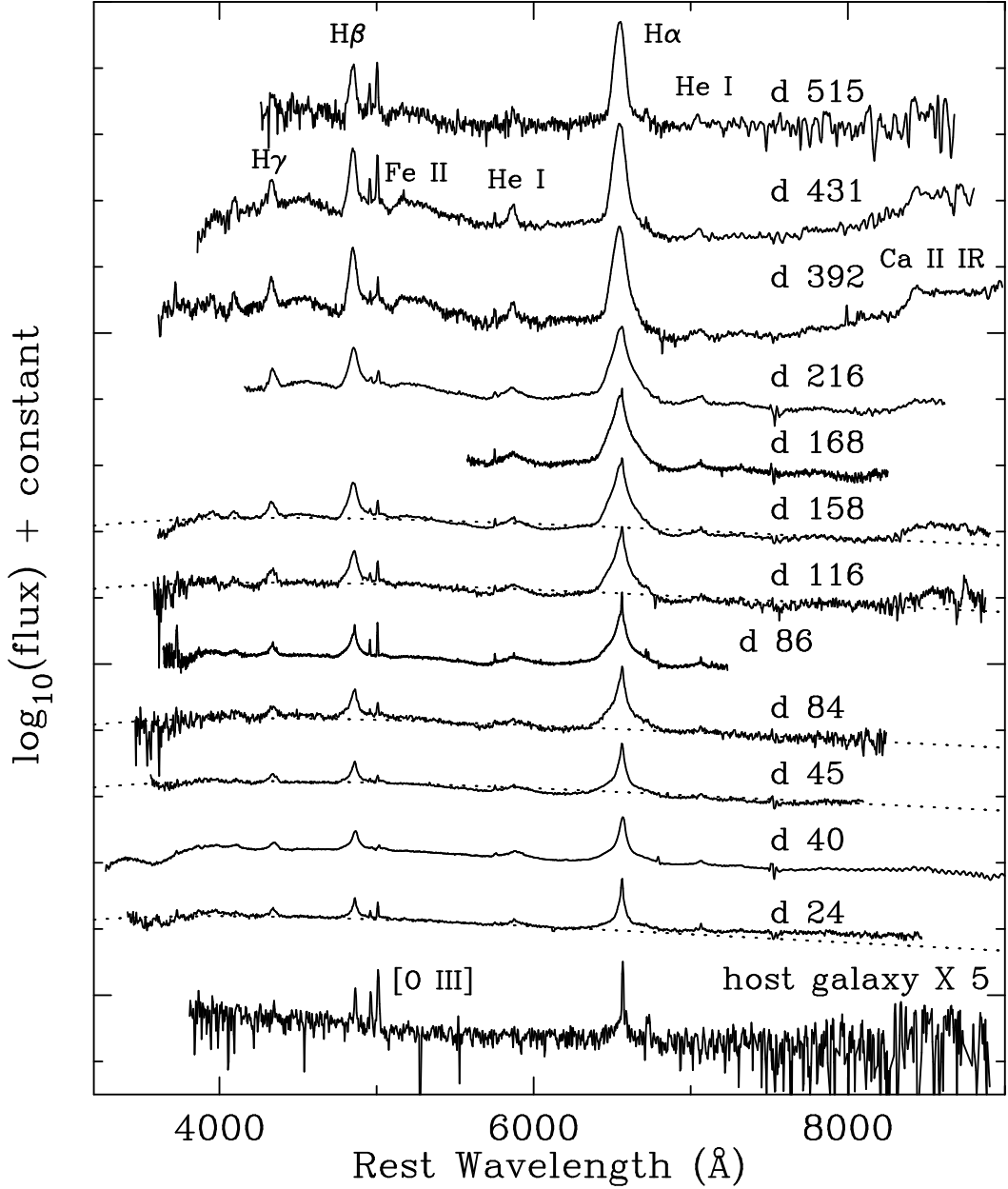}
\caption{Optical spectral evolution of SN 2010jl. The spectra have been corrected for the redshift of the host galaxy ($v_{\rm hel}$ = 3207 km s$^{-1}$) and a reddening of $E(B - V)$ = 0.027 from the Milky Way, and they have been shifted vertically by arbitrary amounts for clarity. The spectra have been normalized to the portion of the red continuum. The dotted curves show representative blackbodies for comparison at the temperatures indicated in Table 5. The numbers on the right-hand side mark the epochs of the spectra in days after $V$-band maximum. The major emission lines are marked.}
\end{figure}

\begin{figure}[ht]
\centering
\includegraphics[angle=0,width=150mm]{f7.ps}
\caption{The spectrum of SN 2010jl at four selected epochs (t $\sim$ 1 month, $\sim$ 3 months, $\sim$ 6 months, and $\sim$ 15 months after $V$ maximum), overplotted with the comparable-phase spectra of SNe 1997cy \citep{tur00}, 1998S \citep{fas01, poz04}, 2006gy \citep{smith10} and 2006tf \citep{smith08}. All spectra shown here have been corrected for the reddening in the Milky Way and redshift of the host galaxy.
The spectra were arbitrarily shifted in the vertical direction for a better display.}
\end{figure}

\begin{figure}[ht]
\centering
\includegraphics[angle=0,width=140mm]{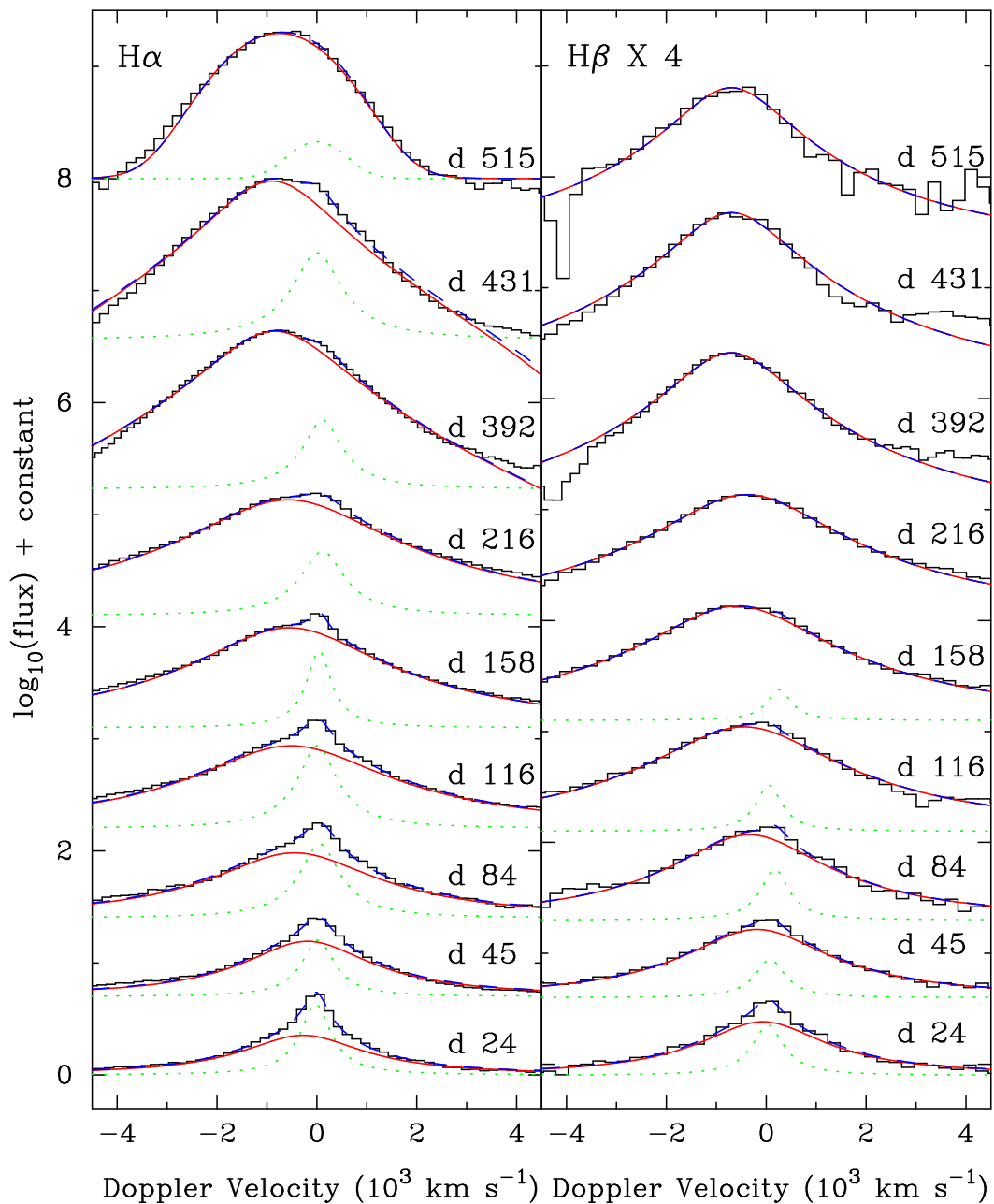}
\caption{{\it Left Panel}: Evolution of the H$\alpha$ line profiles, compared with the double-lorenz/double-gaussian fit. Solid red lines show the velocity distribution of the intermediate-width component on the blue side and dotted green lines show the narrow component on the red side. The blue dashed lines represent best-fit curve to the observed profile. {\it Right Panels}: Similar analysis applied to evolution of the H$\beta$ line profile.}
\end{figure}

\begin{figure}[ht]
\centering
\includegraphics[angle=0,width=140mm]{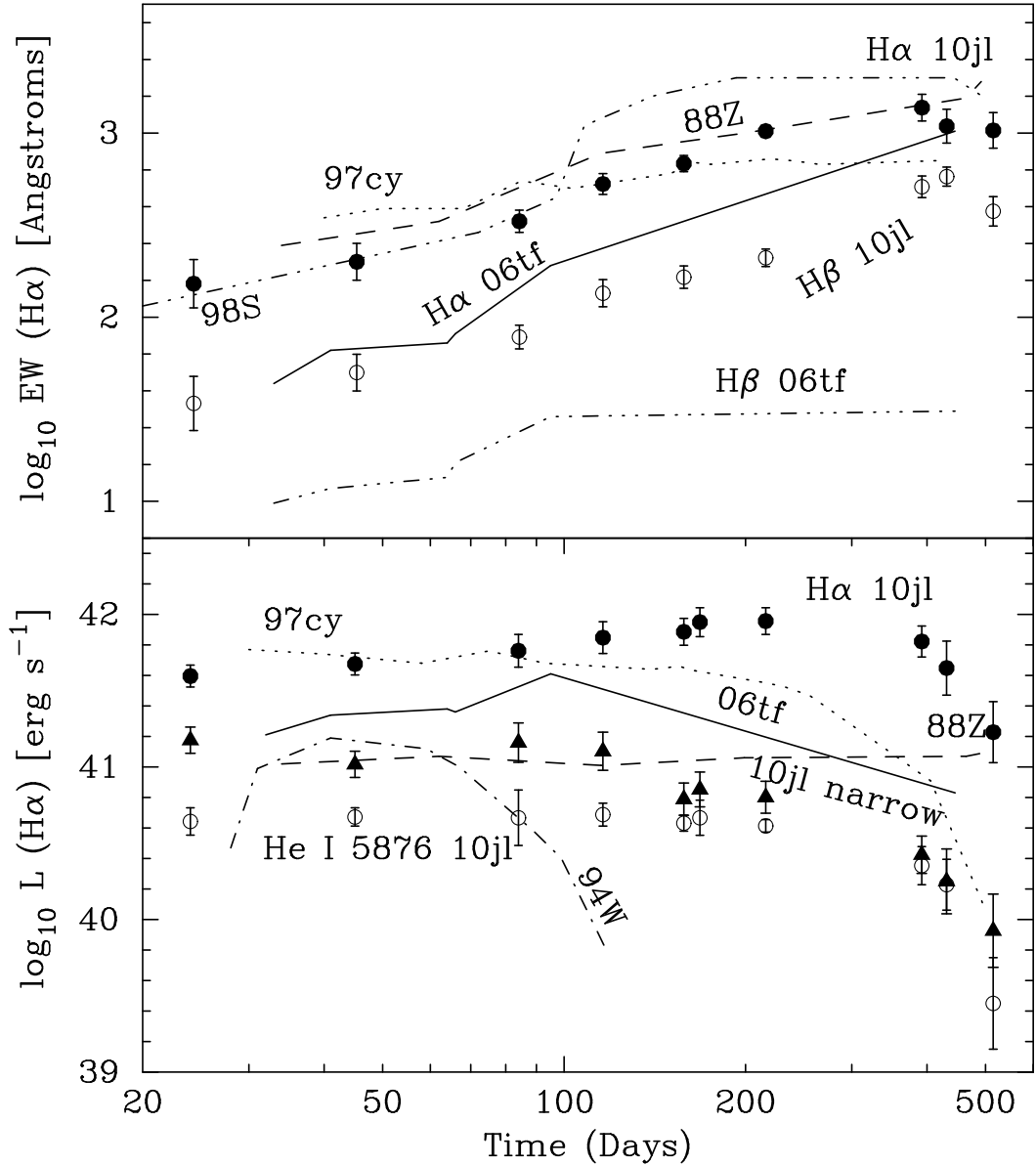}
\caption{{\it Upper Panel}: The Equivalent widths (EWs) of H$\alpha$ and H$\beta$ with time, measured for SN 2010jl. The EW presented here include the contribution of narrow, broad, and intermediate-width components. Overplotted are the evolution of the EW for SNe 1988Z, 1997cy, 1998S \citep{fas00} and 2006tf \citep{smith08}. Note that SNe 1988Z and 1997cy do not have published H$\alpha$ equivalent widths, so borrowed the measurements from Figure 13 of \citet{smith08}. {\it Lower Panel}: Evolution of luminosity of H$\alpha$ lines ($filled$ $circles$) measured in our spectra of SN 2010jl and SNe 1988Z, 1997cy, 1998S \citep{fas00}, and 2006tf \citep{smith08}.}
\end{figure}

\clearpage

\begin{table}
\caption{Magnitudes of Photometric standards in the SN 2010jl Field}
{\small
  \begin{tabular}{lcccccccc}
  \tableline\tableline
Star\tablenotemark{a} & RA.\tablenotemark{b} & Dec. & $U$\tablenotemark{c} & $B$ & $V$ & $R$ & $I$\\
\tableline
1 & 09:43:01.29 & +09:33:08.7 & 16.24(05) & 16.23(02) & 15.58(02) & 15.27(01) & 14.94(01)\\
2 & 09:42:52.33 & +09:32:39.9 & 16.75(07) & 16.39(02) & 15.57(02) & 15.11(01) & 14.64(01)\\
3 & 09:43:09.21 & +09:30:11.8 & 16.26(05) & 15.76(02) & 14.91(01) & 14.43(01) & 13.91(01)\\
4 & 09:43:01.00 & +09:29:49.7 & 14.95(02) & 14.67(01) & 13.94(01) & 13.55(01) & 13.13(01)\\
5 & 09:42:38.44 & +09:29:35.0 & 15.80(04) & 15.59(02) & 14.82(01) & 14.40(01) & 14.00(01)\\
6 & 09:43:05.38 & +09:26:15.9 & 15.72(04) & 15.71(02) & 15.09(01) & 14.76(01) & 14.38(01)\\
7 & 09:42:48.16 & +09:27:17.2 & 16.94(08) & 16.32(02) & 15.39(01) & 14.89(01) & 14.43(01)\\
8 & 09:42:40.93 & +09:26:15.0 & 16.32(05) & 16.22(02) & 15.52(02) & 15.13(01) & 14.79(01)\\
  \tableline
  \end{tabular}
  \tablenotetext{a}{See Figure 1 for a finding chart of SN 2010jl and the reference stars.}
  \tablenotetext{b}{JD 2000}
  \tablenotetext{c}{NOTE.- Uncertainties, in units of 0.01 mag, are $1\sigma$.}
}
\end{table}

\begin{table}
\caption{$UBVRI$ Magnitudes of SN 2010jl from TNT}
{\scriptsize
  \begin{tabular}{lcrccccc}
  \tableline\tableline
  UT Date & JD - 2,450,000 & Phase\tablenotemark{a} & $U$\tablenotemark{b} & $B$ &  $V$ & $R$ & $I$\\
  \tableline
2010 Nov  8 & 5509.36 &  21.36& 13.668(034) & 14.149(024) & 13.736(011) & 13.362(024) & 13.115(024)     \\
2010 Nov  9 & 5510.36 &  22.36& 13.688(034) & 14.161(024) & 13.751(017) & 13.366(024) & 13.132(025)     \\
2010 Nov 11 & 5512.39 &  24.39& 13.706(035) & 14.198(025) & 13.790(024) & 13.358(031) & \nodata         \\
2010 Nov 13 & 5514.34 &  26.34& 13.751(034) & 14.207(026) & 13.779(019) & 13.416(029) & 13.148(028)     \\
2010 Nov 14 & 5515.36 &  27.36& 13.767(030) & 14.238(024) & 13.799(018) & 13.401(024) & 13.154(025)     \\
2010 Nov 15 & 5516.35 &  28.35& 13.760(034) & 14.222(025) & 13.805(018) & 13.430(024) & 13.171(025)     \\
2010 Nov 16 & 5517.34 &  29.34& 13.760(034) & 14.233(025) & 13.803(018) & 13.404(024) & 13.171(024)     \\
2010 Nov 17 & 5518.36 &  30.36& 13.785(037) & 14.256(026) & 13.838(019) & 13.440(024) & 13.173(025)     \\
2010 Nov 21 & 5522.37 &  34.37& 13.816(039) & 14.282(026) & 13.860(021) & 13.471(026) & 13.230(025)     \\
2010 Nov 24 & 5525.32 &  37.32& \nodata     & 14.332(026) & 13.893(018) & 13.505(024) & 13.266(025)     \\
2010 Nov 25 & 5526.42 &  38.42& 13.905(034) & 14.368(025) & 13.920(017) & 13.507(024) & 13.276(025)     \\
2010 Nov 29 & 5530.38 &  42.38& 13.944(034) & 14.391(025) & 13.966(011) & 13.534(024) & 13.297(024)     \\
2010 Dec  2 & 5533.43 &  45.43& 13.982(036) & 14.420(027) & 13.976(018) & 13.564(024) & 13.323(025)     \\
2010 Dec  4 & 5535.40 &  47.40& 14.017(036) & 14.430(027) & 14.002(020) & 13.575(021) & 13.353(026)     \\
2010 Dec  7 & 5538.40 &  50.40& 14.048(034) & 14.455(025) & 14.040(017) & 13.599(024) & 13.377(024)     \\
2010 Dec  8 & 5539.42 &  51.42& 14.054(036) & 14.491(026) & 14.057(018) & 13.610(024) & 13.393(025)     \\
2010 Dec  9 & 5540.38 &  52.38& 14.073(036) & 14.486(026) & 14.076(012) & 13.641(025) & 13.412(026)     \\
2010 Dec 10 & 5541.40 &  53.40& 14.073(034) & 14.511(025) & 14.076(018) & 13.629(024) & 13.409(024)     \\
2010 Dec 11 & 5542.35 &  54.35& 14.133(034) & 14.536(025) & 14.078(017) & 13.643(024) & 13.417(024)     \\
2010 Dec 13 & 5544.42 &  56.42& 14.088(035) & 14.510(025) & 14.077(017) & 13.640(024) & 13.426(024)     \\
2010 Dec 15 & 5546.42 &  58.42& 14.135(035) & 14.532(026) & 14.118(018) & 13.651(025) & 13.444(025)     \\
2010 Dec 23 & 5554.35 &  66.35& \nodata     & 14.587(026) & 14.127(017) & 13.706(024) & 13.484(025)     \\
2010 Dec 31 & 5562.39 &  74.39& 14.311(036) & 14.701(026) & 14.244(018) & 13.776(024) & 13.585(025)     \\
2011 Jan  1 & 5563.36 &  75.36& 14.287(035) & 14.706(026) & 14.254(018) & 13.772(020) & 13.605(025)     \\
2011 Jan  3 & 5565.36 &  77.36& 14.316(031) & 14.719(026) & 14.272(019) & 13.787(025) & 13.605(025)     \\
2011 Jan  4 & 5566.30 &  78.30& \nodata     & 14.734(026) & 14.286(011) & 13.802(024) & 13.600(025)     \\
2011 Jan  5 & 5567.29 &  79.29& \nodata     & 14.739(025) & 14.294(018) & 13.800(024) & 13.624(025)     \\
2011 Jan  9 & 5571.30 &  83.30& 14.336(035) & 14.760(026) & 14.314(018) & 13.821(024) & 13.637(025)     \\
2011 Jan 10 & 5572.26 &  84.26& 14.375(036) & 14.767(026) & 14.331(019) & 13.826(024) & 13.690(025)     \\
2011 Jan 11 & 5573.28 &  85.28& 14.368(034) & 14.772(026) & 14.328(018) & 13.828(024) & 13.668(021)     \\
2011 Jan 12 & 5574.37 &  86.37& 14.363(035) & 14.774(026) & 14.340(019) & 13.839(024) & 13.688(025)     \\
2011 Jan 13 & 5575.27 &  87.27& 14.375(036) & 14.779(021) & 14.334(018) & 13.827(024) & 13.688(025)     \\
2011 Jan 29 & 5591.15 & 103.15& 14.462(036) & 14.857(026) & 14.425(019) & 13.908(024) & 13.778(025)     \\
2011 Jan 30 & 5592.16 & 104.16& 14.487(036) & 14.868(026) & 14.433(019) & 13.913(025) & 13.802(026)     \\
2011 Feb 24 & 5617.14 & 129.14& 14.593(038) & 14.969(026) & 14.540(020) & 13.955(024) & 13.900(026)     \\
2011 Feb 27 & 5620.09 & 132.09& 14.590(045) & 14.995(030) & 14.539(023) & 13.966(027) & 13.898(023)     \\
2011 Feb 28 & 5621.18 & 133.18& 14.568(037) & 14.971(026) & 14.528(020) & 13.967(025) & 13.890(026)     \\
2011 Mar  6 & 5626.99 & 138.99& 14.600(038) & 14.988(026) & 14.548(020) & 13.979(025) & 13.911(026)     \\
2011 Mar  9 & 5630.17 & 142.17& 14.598(037) & 14.980(022) & 14.556(013) & 13.977(026) & 13.933(022)     \\
2011 Mar 22 & 5643.03 & 155.03& 14.642(037) & 15.018(027) & 14.589(020) & 14.009(025) & 13.972(026)     \\
2011 Mar 24 & 5645.06 & 157.06& 14.644(040) & 15.022(026) & 14.604(020) & 14.012(025) & 13.987(026)     \\
2011 Mar 29 & 5650.10 & 162.10& \nodata     & 15.033(027) & 14.612(020) & 14.006(025) & 13.996(026)     \\
2011 Mar 30 & 5651.01 & 163.01& \nodata     & 15.029(029) & 14.639(021) & 14.009(026) & 14.003(028)     \\
2011 Mar 31 & 5652.07 & 164.07& \nodata     & 15.082(032) & 14.621(024) & 13.999(026) & 13.993(028)     \\
2011 Apr  8 & 5660.02 & 172.02& 14.644(040) & 15.042(028) & 14.633(013) & 13.999(026) & 14.029(027)     \\
2011 Apr 10 & 5662.05 & 174.05& 14.653(038) & 15.068(028) & 14.643(021) & 14.027(025) & 14.019(027)     \\
2011 Apr 11 & 5663.06 & 175.06& 14.649(040) & 15.035(028) & 14.642(021) & 14.014(026) & 14.022(027)     \\
2011 May  5 & 5687.02 & 199.02& 14.702(034) & 15.063(023) & 14.666(021) & 14.022(026) & 14.095(029)     \\
2011 Nov 10 & 5876.39 & 388.39& $\sim$15.5 & $\sim$15.9 & $\sim$15.6 & $\sim$14.8 & $\sim$14.8\\
2011 Dec 20 & 5916.36 & 428.36& $\sim$15.7 & $\sim$16.3 & $\sim$15.8 & $\sim$15.1 & $\sim$15.1\\
2012 Mar 13 & 6000.02 & 512.02& $\sim$15.8 & $\sim$16.5 & $\sim$16.1 & $\sim$15.4 & $\sim$15.4\\
  \tableline
  \end{tabular}
  \tablenotetext{a}{Relative to the $V$-band maximum (JD = 2,455,488; \citep{sto11}).}
  \tablenotetext{b}{NOTE.- Uncertainties, in units of 0.001 mag, are $1\sigma$.}
}
\end{table}

\begin{table}
\caption{Journal of Spectroscopic Observations of SN 2010jl}
  \begin{tabular}{lcrlcl}
  \tableline\tableline
  UT Date & JD - 2,450,000 & Phase\tablenotemark{a} & Range(\AA) & Resolution(\AA)\tablenotemark{b} & Instrument\\  \tableline
2010 Nov 11 & 5512.4  & 24.4 & 3500-8500 & 4 & BAO 2.16 m Cassegrain\\
2010 Nov 27 & 5528.4  & 40.4 & 3000-9000 & 3-5 & BAO 2.16 m BFOSC\\
2010 Dec 2 & 5533.4  & 45.4 & 3500-7800 & 4 & BAO 2.16 m Cassegrain\\
2011 Jan 10 & 5572.4  & 84.4 & 3500-8200 & 4 & BAO 2.16 m Cassegrain\\
2011 Jan 12 & 5574.1  & 86.1 & 3600-7300 & 7 & OHP 1.93 m Carelec\\
2011 Feb 11 & 5604.1  & 116.1 & 3600-8800 & 4 & BAO 2.16 m Cassegrain\\
2011 Mar 25 & 5646.1  & 158.1 & 3600-8800 & 4 & BAO 2.16 m Cassegrain\\
2011 Apr 04 & 5656.1  & 168.1 & 5500-8000 & 2 & BAO 2.16 m Cassegrain\\
2011 May 13 & 5704.0  & 216.0 & 4200-8700 & 4 & BAO 2.16 m BFOSC\\
2011 Nov 13 & 5880.4  & 392.4 & 3800-9000 & 3 & YNAO 2.4 m YFOSC\\
2011 Dec 23 & 5919.2  & 431.2 & 3800-9000 & 4 & BAO 2.16 m Cassegrain\\
2012 Mar 16 & 6003.0  & 515.0 & 4300-8700 & 4 & BAO 2.16 m Cassegrain\\
  \tableline
  \end{tabular}
  \tablenotetext{a}{Relative to the $V$-band maximum (JD = 2,455,488; \citet{sto11}).}
  \tablenotetext{b}{Approximate spectral resolution (FWHM intensity).}
\end{table}

\begin{table}
\caption{Spectroscopic Parameters of SN 2010jl}
  \begin{tabular}{r|cc|r@{$\pm$}lr@{$\pm$}l|r@{$\pm$}lr@{$\pm$}l|cc|c}
  \tableline\tableline
Phase\tablenotemark{a} & \multicolumn{2}{|c|}{EW(\AA)} & \multicolumn{4}{|c|}{FHWM(km s$^{-1}$)\tablenotemark{b}} & \multicolumn{4}{|c|}{Blueshift(km s$^{-1}$)\tablenotemark{b}} & H$\alpha$/H$\beta$\\
(day) & H$\alpha$ & H$\beta$ & \multicolumn{2}{|c}{H$\alpha$} & \multicolumn{2}{c|}{H$\beta$} & \multicolumn{2}{|c}{H$\alpha$} & \multicolumn{2}{c|}{H$\beta$} \\
  \tableline
24.4 & -152 & -34 & 2450 & 100 & 2400 & 200 & -276 & 25 & -69 & 78  & 4.5 \\
45.4 & -200 & -50 & 2400 & 100 & 2700 & 100 & -191 & 22 & -185 & 20  & 4.0 \\
84.4 & -332 & -78 & 3000 & 100 & 2650 & 150 & -448 & 32 & -341 & 49  & 4.3 \\
116.1 & -529 & -135 &  3350 & 100 & 3100 & 100 & -520 & 23 & -445 & 22  & 3.9 \\
158.1 & -683 & -165 &  3200 & 100 & 3100 & 100 & -554 & 16 & -594 & 17  & 4.1 \\
216.0 & -1024 & -210 & 3400 & 100 & 3050 & 100 & -576 & 23 & -395 & 13  & 4.9 \\
392.4 & -1375 & -511 & 2600 & 100 & 2250 & 100 & -844 & 17 & -710 & 16  & 2.7 \\
431.2 & -1090 & -580 & 2300 & 100 & 2200 & 100 & -894 & 30 & -694 & 23  & 1.9 \\
515.0 & -1034 & -376 & 2000 & 100 & 2150 & 150 & -753 & 41 & -702 & 31  & 2.8 \\
  \tableline
  \end{tabular}
  \tablenotetext{a}{Relative to the $V$-band maximum (JD = 2,455,488; \citet{sto11}).}
  \tablenotetext{b}{The parameters measured for the intermediate-width component.}
\end{table}

\begin{table}
\caption{Some Physical Properties Derived for SN 2010jl}
  \begin{tabular}{rcccc}
  \tableline\tableline
Phase\tablenotemark{a} & $L_{\rm{Bol}}$ & $T_{\rm{BB}}$ & $\dot{M}_{\rm{CSM}}$ & t$_{\dot{M}}$\\
(day) & (10$^{9}$ $L_{\odot}$) & (K) & ($M_{\odot}$ yr$^{-1}$) & (yr)\\
  \tableline
24 & 9.6 & 7300 & 1.6$\pm$0.3 & -6.5 \\
45 & 7.7 & 6800 & 1.3$\pm$0.3 & -12.2 \\
84 & 5.8 & 6800 & 0.9$\pm$0.2 & -22.6 \\
116 & 4.9 & 6800 & 0.8$\pm$0.2 & -31.2 \\
158 & 4.6 & 6600 & 0.8$\pm$0.2 & -42.8 \\
216 & 4.2 & $\sim$6500 & 0.7$\pm$0.2 & -58.1 \\
388 & 2.0 & $\sim$6500 & 0.3$\pm$0.1 & -104.5 \\
428 & 1.5 & $\sim$6500 & 0.3$\pm$0.1 & -115.3 \\
512 & 1.2 & $\sim$6500 & 0.3$\pm$0.1 & -137.8 \\
  \tableline
  \end{tabular}
  \tablenotetext{a}{Relative to the $V$-band maximum (JD = 2,455,488; \citet{sto11}).}
\end{table}

\end{document}